\documentclass[3p,preprint,12pt,review]{elsarticle}

\usepackage{chemformula}
\usepackage{array,multirow}
\usepackage{graphics}
\usepackage{comment}
\usepackage{wrapfig}
\usepackage{graphicx}
\usepackage{subfig}
\usepackage{amsfonts, amsmath, amsthm, amssymb}
\usepackage{setspace}
\usepackage{float}
\biboptions{sort&compress}
\usepackage{multicol}
\usepackage[pdftex,plainpages=false,pdfpagelabels,colorlinks=true,citecolor=Blue,linkcolor=DarkBlue,urlcolor=DarkRed,filecolor=green,bookmarksopen=true]{hyperref}

\journal{Computational Materials Science}

\begin{document}

\begin{frontmatter}

\title{Structure and Elastic properties of Titanium MXenes: evaluation of COMB3, REAXFF and MEAM force fields}

\author[label1,label2]{Luis F. V. Thomazini}
\author[label2]{Alexandre F. Fonseca}
\ead{afonseca@ifi.unicamp.br}
\address[label1]{Instituto Federal de Educa\c{c}\~ao, Ci\^{e}ncia e Tecnologia de S\~{a}o Paulo, Campus Caraguatatuba, Caraguatatuba, SP CEP 11665‑071, Brazil.}
\address[label2]{Universidade Estadual de Campinas, Instituto de F\'{i}sica Gleb Wataghin, Departamento de F\'{i}sica Aplicada, Campinas, S\~{a}o Paulo, 13083-859, Brazil.}

%
%

\begin{abstract}
Titanium carbide and nitride MXenes are two-dimensional inorganic materials that exhibit noteworthy physical and chemical properties. 
These materials are considered for a variety of technological applications, ranging from energy harvesting to optical and biomedical applications.
Given the growing interest in titanium MXenes, there is an expanding demand for computational studies to predict physical properties and behaviors under diverse physical conditions.  
Complex and large-scale systems necessitate computational methodologies that surpass the constraints imposed by {\it ab initio} calculations. 
In this regard, it is imperative to ascertain the reliability of the computational tools employed to simulate and predict the physical properties of titanium MXenes. 
In this study, the ability of three known classical molecular dynamics (MD) potentials to provide the structural and elastic properties of titanium carbide and nitride MXenes is evaluated. 
The MD potentials that were the focus of this study include the Charge-Optimized Many-Body (COMB3), the Reactive Force Field (REAXFF) and the
Modified Embedded Atom Method (MEAM). 
These three potentials possess two or more sets of parameters, herein referred to as {\it force fields}, capable of simulating Ti-C and Ti-N systems. 
The MD results for the lattice parameter and thickness of the MXenes are then compared to those from DFT calculations found in the literature.
A total of ten force fields were considered; 
of these, two REAXFF and two MEAM ones were identified as the most adequate to simulate both the structure and elastic properties of titanium MXenes. 
Additionally, the values for the linear compressibility of MXenes are presented for the first time. 
Consequently, researchers can utilize the obtained results to design novel MD-based computational studies of titanium MXenes, leveraging the established relative validity of the available force fields.

\end{abstract}

%
%



\end{frontmatter}


\section{Introduction}

\indent

MXenes are two-dimensional (2D) inorganic structures of transition metal carbides and nitrides~\cite{GogotsiACSNANO2019,I1}. 
They are obtained from etching the element ``A'' out of ``MAX'' phases, an abbreviation of the chemical formula  M$_{n+1}$AX$_n$ of ceramic materials, where M is an early transition metal, A is a group IIIA or IVA element, X can be carbon (C) or nitrogen (N), and $n = 1, 2$, or 3~\cite{NaguibADVMATER2014}.
The chemical and structural diversity of MXenes has prompted significant research interest~\cite{I2,I3}, due to their unique optical \cite{I4}, thermal \cite{I5}, mechanical \cite{LipatovSCIADV2018,I6}, and electronic properties \cite{I7}.
First synthesized 14 years ago~\cite{I19,I20,NagubJACS2013}, MXenes continue to be the focus of extensive research, with a wide range of applications, including energy harvesting \cite{I9}, solar cells \cite{I10}, sensors \cite{I11}, electromagnetic shielding \cite{I12}, supercapacitors \cite{I13}, batteries \cite{I14,MaNATCOMM2021}, optical-electronic devices \cite{I15}, composites \cite{I16}, neuromorphic computing~\cite{TeixeiraACSNANO2024}, superlubricity~\cite{LiuJMATERSCI2017,HuangMATERTODAYADV2021}, water filtration \cite{I17} and biomedical treatments \cite{I18}. This explains the perception of MXenes as a novel and impactful 2D entity~\cite{I8}.

Titanium carbides \cite{I19,I20,I21} and nitrides \cite{I22,I23} are members of a known family of titanium MXenes (``M'' = Ti). Together, they have accounted for more than 70 \% of the published works about MXenes~\cite{I24}. 
Their chemical formula is represented by Ti$_{n+1}$X$_n$, where ``X'' can be C or N, as in the ``MAX'' phases. During the process of obtaining titanium carbides and nitrides, some elements such as oxygen (O), fluor (F), chlorine (Cl), or radicals such as hydroxyl (OH) can get bonded to their surfaces. Therefore, their chemical formula is often written as Ti$_{n+1}$X$_n$T$_x$, where T$_x$ is one of the species mentioned above and is called the \textit{termination}~\cite{GogotsiACSNANO2019}. 

In addition to the documented success in MXene synthesis and experimentation, a substantial proportion of titanium MXene studies are computational in nature~\cite{I25}. 
In this regard, first-principle methods have been employed to study the electronic \cite{DFTE1,DFTE3,DFTE4,DFTE6,DFTE2,DFTE7}, optical \cite{DFTO1,DFTO3,DFTO2}, mechanical \cite{PRB2016,DFTM2,DFTM6,DFTM9,DFTE6,DFTM3,I26}, and thermal properties \cite{I26,I28} of titanium MXenes. 
Furthermore, these methods have been used to investigate more special properties and applications such as thermoelectric performance \cite{I27}, electrodes for lithium batteries \cite{DFTE5} and electromechanical actuators~\cite{DFTM7}, to name some examples. 

In addition, computational studies grounded in classical molecular dynamics (MD) have been conducted, encompassing titanium MXenes with and without \emph{terminations}. 
Examples include the thermal performance and stability \cite{I37,I29,I35,I36} of titanium MXenes and the
metal/ion/water intercalations within them \cite{I38,I33,Osti2016ACSAMI}. 
A substantial body of MD research has been dedicated to investigating the mechanical properties of titanium MXenes, including fracture \cite{MDM4,I43,I45,I46,MDMC1}, bending rigidity~\cite{I47}, impact resistance \cite{I48} and friction \cite{I49,I34,I51}. 
The existing literature also encompasses the mechanical characterization of composites~\cite{I42}, porous \cite{I41} and defective \cite{I40} MXenes. 
However, despite these efforts, the full set of mechanical properties for titanium MXenes remains to be fully characterized~\cite{I53}.
Moreover, the dearth of experimental studies on the mechanical properties of individual titanium MXenes hinders the ability to draw meaningful comparisons and conclusions~\cite{LipatovSCIADV2018}. 

The objective of this work is twofold. 
First, the precision of three well-known classical potentials is examined to correctly simulate the structural and elastic properties of Ti$_{n+1}$C$_n$ and Ti$_{n+1}$N$_n$ ($n$ = 1, 2 and 3) MXenes, without terminations, compared to the corresponding DFT data found in the literature. 
In this regard, the MD force fields capable of reproducing these properties are determined. 
The second objective is to determine the values of the following four elastic mechanical properties of all titanium MXenes, Young's modulus, $E$, linear compressibility, $\beta$, Poisson's ratio, $\nu$, and shear modulus, $G$. 
This work will be the first to predict the values of $\beta$ for all titanium MXenes. 

The classical potentials that will be studied here are the 3rd generation of the Charge-Optimized Many-Body (COMB3)~\cite{I54}, the Reactive Force Field (REAXFF)~\cite{I55} and the Modified Embedded Atom Method (MEAM)~\cite{I56}. 
It is noteworthy that both REAXFF and COMB3 possess  more than one set of parameters to simulate systems formed by Ti, C and/or N. 
Consequently, each set will be tested. 
Each set of parameters is designated as a \emph{force field} and will be named by the initial of the potential abbreviation and a number. 
In the case of REAXFF, the force fields named ``R1''~\cite{R1}, ``R2''~\cite{I33}, ``R3''~\cite{R3}, ``R4''~\cite{R4}, ``R5''~\cite{R5} and ``R6''~\cite{R6} have parameters developed to simulate the interactions between Ti and C. 
The first five REAXFF force fields also describe the interaction between Ti and N. 
In the case of COMB3, there are two force fields capable of simulating Ti, C and N.
One of them, named ``C1'', was developed to simulate more general Ti/TiC interfaces~\cite{LiangJPCC2016}, and the other, named ``C2'', was developed to simulate titanium-graphene interfaces~\cite{FonsecaACSAMI2017}. 
In the context of MEAM, two force fields, named ``M1'' and ``M2'', developed by Kim and Lee~\cite{KimACTAMAT2008} to simulate bulk TiC and TiN, respectively, were considered. There are nine (eight) force fields that can simulate Ti and C (Ti and N) to be tested. 

The present study evaluated the performance of three classical potentials and their corresponding force fields in the acquisition of the structural and mechanical properties of six titanium MXenes, three carbides and three nitrides. 
Evaluation of the force fields reveals that only the R4 (R3) force field is shown to be capable of simulating the Ti$_2$C (Ti$_2$N) MXene. 
In contrast, both M1 and M2 force fields exhibited the strongest agreement with \emph{ab initio} data for Ti$_3$C$_2$, Ti$_4$C$_3$, Ti$_3$N$_2$, and Ti$_4$N$_3$ MXenes.  
The remaining force fields from REAXFF or COMB3 potentials demonstrate a partial capacity to simulate the structure or the elastic constants of (though not all) titanium MXenes. 
However, certain force fields have been demonstrated to be entirely inadequate in describing the atomic structure of some MXenes. 
These results underscore the persistent challenge of accurately modeling MXenes, as they necessitate interatomic potentials capable of ensuring structural fidelity while reliably predicting mechanical behavior.

In Section \ref{sec2}, the MXene structures, the theoretical background to obtain the elastic constants, and the computational methods used to simulate them are described. 
Furthermore, the criteria for determining the force fields suitable for the simulation of some or all titanium MXenes are presented. 
In Section \ref{sec3}, the results of the tests are presented, and a list of the most effective force fields is provided.  
In Section \ref{sec4}, the most effective force fields are employed to ascertain the four elastic properties of MXenes. 
Finally, Section \ref{sec5} presents a summary of the primary results and conclusions. 

\section{Structures, Elastic Theory and Computational Methods}
\label{sec2}

This Section presents the structures of the titanium MXenes that are the subject of this study, the theory employed to obtain the elastic constants of the MXene, and the computational methods employed in the simulations.

\subsection{\texorpdfstring{Structure of Ti$_{n+1}$X$_n$}\mbox{} (X = C or N), MXenes}
\label{sec21}

\begin{figure}[H]
\renewcommand{\figurename}{Figura}
\centering
\includegraphics[scale=0.35]{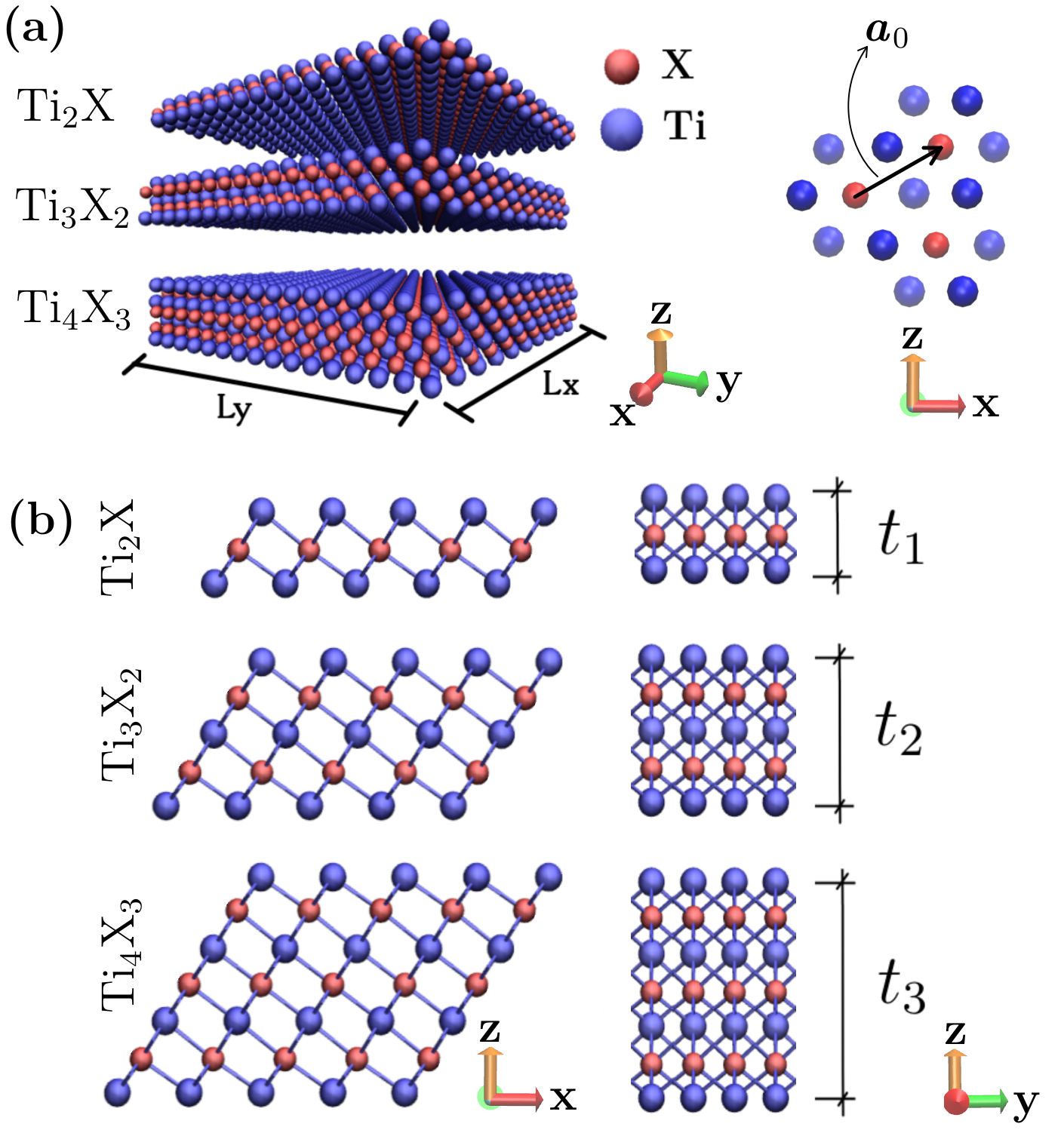}
\caption{Atomic structures of the Ti$_{n+1}$X$_n$ (X = C or N) MXenes. (a) Perspective views of Ti$_2$X, Ti$_3$X$_2$, Ti$_4$X$_3$ (X = C or N) of $L_x$ and $L_y$ dimensions, and the in-plane lattice parameter. 
(b) Lateral views of the structures, including the definition of the thickness, $t_n$.}
\label{S1}
\end{figure}

\indent The formula for the MXenes that will be considered here is as follows: Ti$_{n+1}$X$_n$, i.e., that is to say, without \emph{terminations}. 
In essence, these structures are two-dimensional (2D) and can be regarded as comprising titanium (Ti) layers intercalated by carbon or nitrogen, ``X'', layers. 
As illustrated in Figure \ref{S1}(a),  perspective views of the crystalline structure of Ti$_2$X, Ti$_3$X$_2$ and Ti$_4$X$_3$ MXenes are presented.
The planar lattice parameter, designated as $a_0$, is also displayed. 
Figure \ref{S1}(b) presents the lateral views of the Ti$_{n+1}$X$_n$ MXenes (their zx and zy-planes) and indicates their atomic layers and thicknesses, $t_n$. 

Samples of a square shape with side lengths given by $L_x \approx L_y \approx 50$ \AA, were defined for the simulations of all MXenes. 
The Ti$_{n+1}$X$_n$ (X = C or N) MXenes examined in this study contain 960 ($n=1$), 1600 ($n=2$) and 2240 ($n=3$) atoms, respectively. 
The armchair and zigzag directions were selected to align with the x- and y-axes of all MXenes, respectively (Figure \ref{S1}a).

\subsection{The elastic constants and the four elastic properties}
\label{sec22}

\indent The elastic energy, $U$, per unit of area, accumulated under the deformation of 2D anisotropic structures, can be expressed through the elastic constants $C_{11}$, $C_{12}$, $C_{22}$ and $C_{66}$, according to the following equation \cite{2D1,2D2}, 
\begin{equation}
U = U_0 + \frac{1}{2}C_{11}{\varepsilon^2_{xx}} + \frac{1}{2}C_{22}{\varepsilon^2_{yy}} + C_{12}\varepsilon_{xx}\varepsilon_{yy} + 2C_{66}{\varepsilon^2_{xy}} \, ,
\label{1}
\end{equation}
where $U_0$ is the energy of the optimized structure, $\varepsilon_{xx}$ and $\varepsilon_{yy}$ represent the strains along the x and y directions, respectively, and $\varepsilon_{xy}$ denotes the shear strain. 

The following 2D elastic mechanical quantities can be directly evaluated from the elastic constants, $C_{ij}$, through~\cite{2D3,Kanegae2022CarbonTrends}:

\begin{equation} 
\begin{aligned}
    E_x &= \frac{C_{11}C_{22} - C_{12}^2}{C_{22}}, & \quad E_y &= \frac{C_{11}C_{22} - C_{12}^2}{C_{11}} \\[10pt]
    \beta_x &= \frac{C_{22} - C_{12}}{C_{11}C_{22} - C_{12}^2}, & \quad \beta_y &= \frac{C_{11} - C_{12}}{C_{11}C_{22} - C_{12}^2} \\[10pt]
    \nu_{xy} &= \frac{C_{12}}{C_{22}}, & \quad \nu_{yx} &= \frac{C_{12}}{C_{11}}, &  G &= C_{66} \, , 
\end{aligned}
\label{2}
\end{equation}
where $E_x$, $E_y$, $\beta_x$ and $\beta_y$ denote the Young’s modulus, $E$, and the linear compressibility, $\beta$, of the structures in the x and y directions, respectively. 
The Poisson's ratios, $\nu _{xy}$ and $\nu _{yx}$ are defined as the ratios of the areas under the stress-strain curves when the load is applied along the x and y directions, respectively. 
Finally, $G$ is the shear modulus of the structures. 

It is noteworthy that the $C_{66}$ elastic constant can be derived from the other $C_{ij}$ values through the following relationship~\cite{2D1}:
\begin{equation} 
    C_{66}=\frac{1}{4}\left(C_{11}-2C_{12}+C_{22}\right) \, .
\label{C66}
\end{equation}

\subsection{Computational methods}
\label{sec23}

\indent All classical molecular dynamics (MD) simulations are performed using the LAMMPS computational package \cite{CM1}. 
The first objective of this work is to evaluate the ability of the force fields to simulate the structure and mechanical properties of titanium MXenes. 
To this end, protocols of MD simulations were designed to obtain the optimized structure and the elastic constants, $C_{ij}$, of all titanium MXenes. 
These protocols are derived from those developed by Kanegae and Fonseca~\cite{Kanegae2022CarbonTrends} to study the elastic properties of graphyne structures. 
The second objective of this work, which is to calculate the elastic properties of all titanium MXenes, does not require computational simulations.  
The protocols of MD simulations are detailed in the following subsections. 

The MD simulations were performed using the COMB3, REAXFF, and MEAM potentials. As previously mentioned in the Introduction, a total of six force fields were considered for REAXFF, with five of them pertaining to Ti, C, and N (R1 \cite{R1}, R2 \cite{I33}, R3 \cite{R3}, R4 \cite{R4}, and R5 \cite{R5}), and one specifically addressing Ti and C (R6 \cite{R6}). 
For COMB3, two force field files were utilized, both of which were capable of describing Ti, C, and N (C1 \cite{LiangJPCC2016} and C2 \cite{FonsecaACSAMI2017}). 
For the MEAM, two force fields were utilized: one for the Ti–C systems (M1) and another for the Ti–N systems (M2) \cite{KimACTAMAT2008}. 

Unless otherwise indicated, the simulations are performed under periodic boundary conditions (PBC) within the xy-plane. 
The x- and y-dimensions of the structures, denoted $L_x$ and $L_y$, will be referred to as ``box sizes''. 

\subsubsection{Protocols of MD Simulations}
\label{sec231}

\indent The MD protocols employed in this work to evaluate the force fields are described in this section. 
To enhance comprehension, the explanations of the protocols are divided into two distinct processes.
The objective of these protocols is to enable the acquisition of the $C_{ij}$ of all titanium MXene structures. 

The first process, designated as ``\textit{minimization}'', is employed to ascertain the structure that exhibits the lowest cohesive potential energy. 
This process entails the optimization of the box sizes that delineate the structure. 
The objective is accomplished through a series of iterative minimization cycles, the particulars of which will be expounded upon subsequently. 
Upon completion of this process, the set of atomic coordinates of the structure will be designated as the ``{\it optimized structure}''.

\begin{figure}[H]
\centering
\includegraphics[scale=0.4]{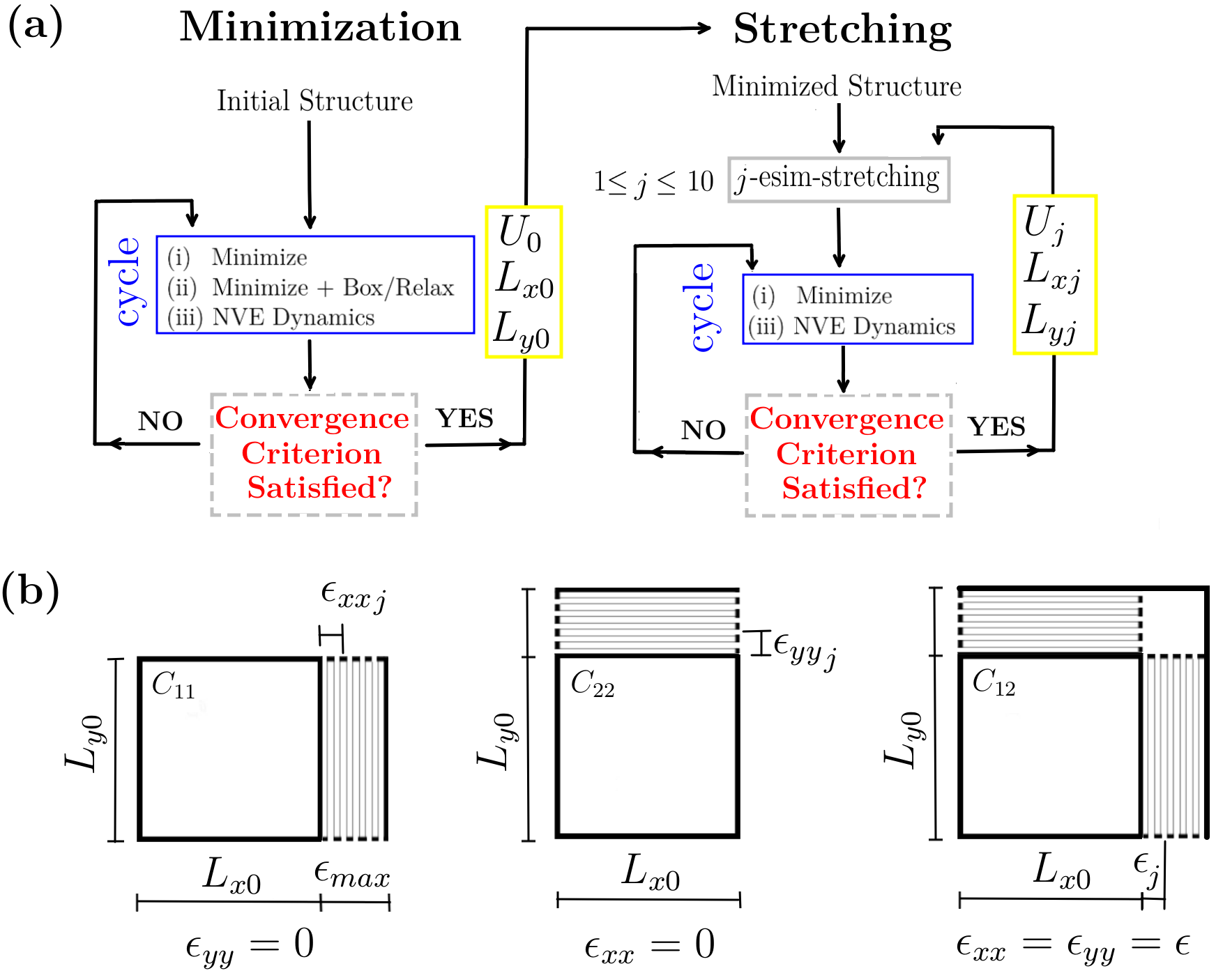}
\caption{Schematic representation of the simulation protocols used for calculating the optimized structure and its elastic constants $C_{ij}$. (a) \textit{Minimization} and {\it stretching} processes. (b) Illustrations of how the tensile strain is applied in the {\it stretching} part of the protocol and the corresponding $C_{ij}$ that can be obtained from them. 
The details are explained in the text. 
}
\label{CM}
\end{figure}

The second process, designated as ``\textit{stretching}'', consists of the implementation of a sequence of energy minimizations of tensile strained structures, starting from the optimized one. 
The simulations yield a set of points representing the energy versus strain relationship, which is necessary for calculating the $C_{ij}$. 
Figure \ref{CM}(a) presents the schemes of both parts of the protocols.

The {\it minimization} part of the protocol entails the execution of cycles comprising three LAMMPS' MD algorithms or processes: (i) energy minimization; (ii) energy minimization with box relaxation; and (iii) a brief dynamical simulation of the system with a constant number of atoms (N), volume (V) and total energy (E). 
The first process involves adjusting the atom positions in accordance with the energy minimization LAMMPS' algorithms, while maintaining constant box sizes. 
The subsequent step involves a concurrent modification of the box dimensions in the x and y directions. 
In this context, ``box relaxation'' signifies that the energy minimization algorithm not only searches for the optimal atom positions that minimize the total energy of the structure but also looks for the sizes of the x- and y-dimensions, under PBC, of the system that minimize the total energy. 

The third NVE dynamical process is  necessary for the verification of the convergence of the preceding steps. 
In the NVE dynamical process, atoms begin to move and acquire kinetic energy if their positions do not correspond precisely to a real local minimum of the energy. 
In the event of such an occurrence, it can be deduced that the preceding two processes of energy minimization were inadequate in identifying the optimal atom coordinates and box sizes that truly minimize the total energy of the structure. 
It was observed that the greater the kinetic energy acquired by the system after a few timesteps of this NVE dynamic simulation, the further the system was from the local or global energy minimum after the minimization processes. 
The subsequent criterion is then utilized to identify the smallest energy structure.
In the event that the kinetic energy accumulated following approximately 1300 timesteps of the NVE dynamics is less than 10$^{-6}$ (in energy units), the simulation is terminated. 
Otherwise, the sequence of processes (i), (ii) and (iii) is reiterated.
The selection of 10$^{-6}$ was made after the execution of tests with various structures and force fields at smaller values. 
It was observed that their structural properties remained constant or exhibited changes at the third or higher decimal places.

Upon completion of this stage, the simulation provides the {\it optimized structure} with dimensions $L_{x0}$ and $L_{y0}$, lattice-parameter $a_0$, and the value of its initial potential energy, $U_0$. 

The rationale for executing cycles of this set of ``three processes'' stems from two key considerations.
Firstly, it is imperative to execute dynamic tests~\cite{Sihn2015CARBON} to ascertain that the structure attains the minimum energy level to the extent desired. 
Secondly, the recognition that the algorithms employed for energy minimization are ill-defined when used in conjunction with the box size relaxation (see LAMMPS' manual~\cite{CM1}).

The \textit{stretching} process consists of the implementation of a series of 10 tensile strains on the optimized system, which is derived from the preceding {\it minimization} process. 
The conceptual framework of this process is illustrated in the right panel of Figure \ref{CM}(a). 
Each tensile-strain step is labeled by $j$, $1\leq j\leq10$, and is designated as a ``$j$-esim-stretching'' step. 
In each of these steps, the simulation box undergoes a tensile strain deformation of 0.1$\%$, applied in one or both directions, contingent on the type of $C_{ij}$ being calculated (refer to Figure \ref{CM}(b) and subsequent subsections \ref{sec232} and \ref{sec233}). 
Subsequently, cycles of the (i) and (iii) processes described in the {\it minimization} protocol are implemented to ascertain the optimized tensile-strained structure and its potential energy. 
The step (ii), which permits box relaxation, is not applied because the structure is expected to remain strained and with its box sizes fixed. 

The same criterion of convergence of the kinetic energy of the structure, which was utilized in the {\it minimization} process, is employed for each $j$-esim-stretching step. 
Upon satisfying this criterion, the $(j+1)$-esim tensile strain step of this protocol commences by subjecting the previous structure to an additional 0.1 $\%$ strain. 
This  process is repeated until $j=10$ or the total strain is 1 \%.
A set of ten values of $L_{xj}$, $L_{yj}$ and the corresponding converged potential energy, $U_j$, in conjunction with $L_{x0}$, $L_{y0}$, and $U_0$, will constitute a dataset from which one of the $C_{ij}$ elastic constants can be derived.

For the COMB3 and REAXFF potentials, the corresponding charge equilibration (QEQ) algorithms were kept activated to perform charge equilibration of the system at each time step, with a convergence criterion of 10$^{-6}$. 
MEAM does not require charge equilibration. 
All LAMMPS' energy minimization calculations were performed with stopping energy and force tolerances set to 10$^{-10}$ and 10$^{-10}$ eV/\AA \, respectively, for COMB3 and MEAM. For REAXFF, the energy and force tolerances were the same, except for the force was given in kcal/mol/\AA. 
The timestep for all NVE dynamical simulations was set in 0.1 fs. 

The protocols to obtain $C_{ii}$, $i=1,2$, and $C_{12}$ differ in how tensile strain is applied to the structure during the main {\it stretching} protocol. 
These protocols are illustrated in Figure~\ref{CM}(b) and explained in the following subsections.

\subsubsection{$C_{ii}$, $i=1,2$, MD Protocols}
\label{sec232}

\indent To obtain $C_{11}$ or $C_{22}$, uniaxial tensile strains should be applied to the structure in the x or y direction, respectively.
The simulation steps in the {\it stretching} protocol should be performed by applying strains in only one direction while fixing the size of the structure in the transverse direction.  
For example, to obtain $C_{11}$, strains are applied only along the x-direction while keeping the box size fixed along the y-direction. 
This means that $\varepsilon_{yy}=\varepsilon_{xy}=0$ during the set of simulations in the stretching protocol. 
For this protocol, the equation for the elastic energy of the structure is:
\begin{equation}
U = U_0 + \frac{1}{2}C_{11}{\varepsilon^2_{xx}} \, ,
\label{C11}
\end{equation}
where $U_0$ is the energy of the fully optimized structure, as obtained from the {\it minimization} protocol. 
A tensile strain scheme illustrating the above process is shown in the left panel of Figure \ref{CM}(b). 

The process to obtain $C_{22}$ is the same, except $\varepsilon_{xx}$ and $\varepsilon_{xy}$ are set to 0 and the strains are applied only along the y-direction.
The energy equation becomes similar to the equation (\ref{C11}), except $C_{11}$ and $\varepsilon_{xx}$ are replaced by $C_{22}$ and $\varepsilon_{yy}$. The corresponding scheme is shown in the middle panel of Figure \ref{CM}(b). 

Then, $C_{11}$ and $C_{22}$ can be obtained by fitting the corresponding energy-versus-strain curves with a parabola.

\subsubsection{$C_{12}$ MD Protocol}
\label{sec233}

\indent To obtain $C_{12}$, the {\it stretching} process must be performed under a biaxial tensile strain.
This means that the same amount of tensile strain must be applied in both the x- and y-directions. 
In other words, 
$\varepsilon_{xx}=\varepsilon_{yy}\equiv\varepsilon$ and $\varepsilon_{xy}=0$. Equation (\ref{1}) becomes:
\begin{equation}
U = U_0 + \frac{1}{2}M\varepsilon^2 \, ,
\label{C12}
\end{equation}
where $M$ is simply given by
\begin{equation}
M = \frac{1}{2}(C_{11}+2C_{12}+C_{22}) \, .
\label{M}
\end{equation}

\noindent After running the $C_{ii}$ MD protocols to obtain $C_{11}$ and $C_{22}$, the value of $C_{12}$ can be found using equation (\ref{M}). 

\subsection{Tests and criteria to evaluate force fields}
\label{sec24}

The primary objective of this study is to assess the efficacy of MD force fields in describing the structural characteristics and elastic properties of titanium MXenes. 
To this end, three tests are hereby proposed. 
Two tests are employed to evaluate the force field's capacity to simulate the structural properties of titanium MXenes, and one test assesses its ability to simulate elastic properties. The tests are applied sequentially. 
The initial evaluation is of a qualitative nature, while the subsequent two are quantitative. 
The initial two tests are instrumental in the identification and elimination of inadequate or unfavorable force fields. 
The final test is of a classificatory nature. 
The subsequent subsections offer comprehensive explanations of the evaluation criteria.

\subsubsection{First test}    
\label{1stcriterion}

The first evaluation test, of a qualitative nature, involves the verification of whether the force field employed in the {\it minimization} protocol is indeed capable of optimizing the MXene structure. 
The basis of this criterion is a visual inspection of the regularity of the structure or an inspection of a given structural variable, such as the lattice parameter. 
For instance, if the structure is deformed, broken, amorphous, or if the force field generates any kind of computational or numeric error, the corresponding force field is considered inadequate and discarded. 
In the event that the minimization protocol is executed for a specific structure with a designated force field, and the resultant structure exhibits a contradictory outcome, such as a substantial variation in the lattice parameter values, the force field is deemed inadequate for simulating the corresponding structure, according to this criterion. 

Each force field is subjected to testing for each MXene, for which there are simulation parameters available. 
In the event that a given force field is found to be ineffective in minimizing one MXene, yet is capable of achieving this outcome for other MXenes, it will be further considered for the latter.

The force fields for which the optimized MXene structures retain the planar crystalline form will be considered for the second, quantitative, criterion.

\subsubsection{Second test}
\label{2ndcriterion}

The objective of the second test is to verify the agreement, within a certain tolerance, between density functional theory (DFT) calculations found in the literature and those from the force fields that satisfied the first test, of two quantities: the lattice parameter, $a_0$, and the thickness, $t$, of the Ti$_{n+1}$C$_n$ MXenes. 
Table \ref{TSCL} presents the values of these quantities obtained from DFT calculations~\cite{DFTE1,DFTE2,DFTE3,DFTE4,DFTE5,DFTE6,DFTE7,DFTE6,DFTM2,DFTM7,DFTM9,DFTO1,DFTO2,DFTO3}.
\begin{table}[h!]
\caption{Lattice parameter, $a_0^{\mbox{\tiny{DFT}}}$ and thickness, $t^{\mbox{\tiny{DFT}}}$, of titanium carbide and nitride MXenes, in \AA, from DFT calculations found in the literature. The minimum and maximum values are indicated in {\bf bold} font.}
\centering
\renewcommand{\arraystretch}{1.3} 
\setlength{\tabcolsep}{25pt} 
\resizebox{\textwidth}{!}{%
\small
\begin{tabular}{c c c}
\hline \hline
 \textbf{Structure} & $a_0^{\mbox{\tiny{DFT}}}$    & $t^{\mbox{\tiny{DFT}}}$  \\
\hline \hline
 \multirow{2}{*}{\centering Ti$_2$C}  
                            &  {\bf 3.00}\cite{DFTE4}   & {\bf 2.23}\cite{DFTO1}, 2.29\cite{DFTE4,DFTE2, DFTE6}
                           \\  
                           & {\bf 3.04}\cite{DFTE1,DFTE3,DFTE2,DFTE6,DFTM2,DFTM9,DFTO1}  & {\bf 2.31}\cite{DFTE6,DFTM2,DFTO3,DFTM7}
                           \\  
\hline                  
\multirow{3}{*}{\centering Ti$_3$C$_2$}  
                         &  {\bf  2.91}\cite{DFTO1}, 3.07\cite{DFTE4} & 
                         {\bf 4.60}\cite{DFTE4}, 4.63\cite{DFTO3} \\
                         &   3.08\cite{DFTE1,DFTE3}, 3.09\cite{DFTE2} &  4.64\cite{DFTE6}        \\
                         &  {\bf 3.10}\cite{DFTE6,DFTO3} &  4.66\cite{DFTE2} {\bf 4.69}\cite{DFTO1} \\
\hline
\multirow{2}{*}{\centering Ti$_4$C$_3$}  
                         &  {\bf 3.07}\cite{DFTE4} & {\bf 7.14}\cite{DFTE4}, 7.15\cite{DFTE6}
                         \\
                         &  {\bf 3.09}\cite{DFTE1,DFTE5,DFTE6} &  {\bf 7.16}\cite{DFTE5}
                         \\
\hline \hline
\multirow{2}{*}{\centering Ti$_2$N } 
                         &   {\bf 2.98}\cite{DFTE1,DFTE6,DFTM9} & {\bf 2.27}\cite{DFTO1}, 2.29\cite{DFTE6}
                         \\
                         & {\bf 3.03}\cite{DFTO1} & {\bf 2.31}\cite{DFTE7}
                         \\
\hline
\multirow{2}{*}{\centering Ti$_3$N$_2$ }  
                         & {\bf 3.00}\cite{DFTE1}, 3.01\cite{DFTE7} & {\bf 4.60}\cite{DFTO1}, 4.73\cite{DFTE6}
                         \\
                         &   3.05\cite{DFTE6}, 3.06\cite{DFTO1},
                         {\bf 3.07}\cite{DFTO2} & {\bf 5.12}\cite{DFTO2}
                         \\
\hline
\multirow{1}{*}{\centering Ti$_4$N$_3$ }  
                         &   {\bf 2.99}\cite{DFTE1,DFTE6}  & {\bf 7.36}\cite{DFTE6}
                         \\
\hline \hline
\end{tabular}%
}
\label{TSCL}
\end{table}

As illustrated in Table \ref{TSCL}, the majority of structures exhibit multiple DFT values for either the lattice parameter, $a_0$, or thickness, $t$. 
In the context of this particular test, each distinct DFT value reported in the literature is considered equivalent to another. 
For each structure, distinct DFT values delineate an interval. 
The criterion of acceptance of a force field is, therefore, defined as the discrepancy between the optimized titanium MXene structure simulated with the force field and one of the extremes of the interval of DFT values. 
The criterion for the force field is considered met if the MD values of $a_0$ and $t$ fall within the interval of DFT values or deviate by no more than 5 \% from the extremes of the range of DFT values. 
Conversely, the force field is regarded as inadequate for simulating the titanium MXene system.

An additional criterion that is verified concerns the thickness of the structure. 
If the thickness of the optimized structure exceeds 33.3 \% of that derived from DFT calculations, the corresponding force field is deemed inadequate.

The force fields that yield optimized structures that satisfy the $a_0$ and $t$ criteria will undergo the next and final test.

\subsubsection{Third test}
\label{3rdcriterion}

The final evaluation of the force fields to simulate titanium MXenes is based on the accuracy with which they provide the values of the elastic constants, $C_{ij}$ in comparison to DFT calculations~\cite{DFTE2,DFTE4,DFTE6,DFTO3,DFTM2,DFTM3,DFTM9,PRB2016}. 
The values are presented in Table \ref{TECL}.
The $C_{ij}$ will be calculated for all force fields that satisfy the previously outlined criteria.
Following this, a comparison is made between the MD results and the DFT values. 
This comparison is employed to establish a classification of the force fields based on their capacity to predict the structure and mechanical properties of titanium MXenes.

\begin{table}[h!]
\caption{Elastic constants, $C_{ij}^{\mbox{\tiny{DFT}}}$, of titanium MXenes,  in N/m, from DFT calculations found in the literature. The minimum and maximum values of $C_{11}^{\mbox{\tiny{DFT}}}$ are indicated in {\bf bold} font.}
\centering
\renewcommand{\arraystretch}{1.3} 
\setlength{\tabcolsep}{15pt} 
\small 
\begin{tabular}{ c c c c c }
\hline \hline
 \textbf{Structure} & $C_{11}^{\mbox{\tiny{DFT}}}$  & $C_{12}^{\mbox{\tiny{DFT}}}$  & $C_{22}^{\mbox{\tiny{DFT}}}$  & $C_{66}^{\mbox{\tiny{DFT}}}$   \\
\hline \hline
\\
 \multirow{3}{*}{\centering Ti$_2$C}
                & {\bf 130}\cite{DFTE2}, 135\cite{DFTM9} &  &  &
                          \\
                & 137\cite{DFTM2}   139\cite{DFTE6}  140\cite{DFTO3} & 32\cite{DFTM2}  35\cite{DFTM3} &  153\cite{DFTM3} & 54\cite{DFTM3}
                                \\
                &   146\cite{DFTE4} {\bf 151}\cite{DFTM3} &  37\cite{DFTM9} & &  \\
                           \\
\hline
\\
\multirow{3}{*}{\centering Ti$_3$C$_2$} 
                & {\bf 219}\cite{DFTE6}  227\cite{DFTO3} &  
                                \\
                &   231\cite{DFTE2}   241\cite{DFTE4,PRB2016} & 40\cite{DFTM3}  &  257\cite{DFTM3}    & 105\cite{DFTM3}
                                \\
                & {\bf 253}\cite{DFTM3}    \\
                           \\
\hline           
\\
\multirow{3}{*}{\centering Ti$_4$C$_3$}  
                & {\bf 312}\cite{DFTM3} & 
                                \\
                &   328\cite{DFTE6} & 49\cite{DFTM3}  &  306\cite{DFTM3}  & 132\cite{DFTM3}
                                \\
                &  {\bf 366}\cite{DFTE4}   &      \\
\\
\hline \hline
\\
\multirow{2}{*}{\centering Ti$_2$N } 
                & {\bf 150}\cite{DFTE6}
                                \\
                & {\bf 154}\cite{DFTM9} & 41\cite{DFTM9} & -- & -- 
                                \\
\\                                
\hline   
\\
\multirow{1}{*}{\centering Ti$_3$N$_2$ }  
                &   263\cite{DFTE6} & -- & -- & --
                                \\
\\                                
\hline 
\\
\multirow{1}{*}{\centering Ti$_4$N$_3$ }  
                &   369\cite{DFTE6}  & -- & -- & --
                                \\
\\ \hline \hline
\end{tabular}
\label{TECL}
\end{table}

Two supplementary tests can be implemented to verify the integrity of the results and the adequacy of the corresponding force field to simulate titanium MXenes.
One such test involves verifying the relationship between $C_{11}$ and $C_{22}$. 
The {\it stretching} protocols have been meticulously designed to ensure the independent acquisition of $C_{11}$ and $C_{22}$. 
In accordance with the square symmetry of the titanium MXene structures, it is anticipated that $C_{11}=C_{22}$. 
In the event of a substantial discrepancy between $C_{11}$ and $C_{22}$, the test will be deemed unsuccessful, and the corresponding force field will be disregarded.

The second test is concerned with the structural integrity of the structure. 
The set of $C_{ij}$ values obtained from the MD simulations must satisfy the criteria for the strain energy be always positive~\cite{Marcin20192DMater}:
\begin{equation}
C_{22}=C_{11} > 0 , \quad C_{33} > 0 , \quad \mbox{and} \quad  C_{11} > |C_{12}| \, . 
\label{conditionsCij}
\end{equation}

As can be seen in Table \ref{TECL}, there is a scarcity of DFT values for $C_{ij}\neq C_{11}$. 
Consequently, the quality of the force fields to be used to simulate the mechanical properties of titanium MXenes will be evaluated primarily based on the values of $C_{11}$. 
Additionally, a conspicuous discrepancy in the DFT values of $C_{11}$ across different references is noteworthy. 
The most substantial percentage discrepancy in $C_{11}$ values is approximately 17.3 \% for Ti$_4$C$_3$. 
It is reasonable to expect that the force fields will not produce precise results for all $C_{ij}$ when compared to DFT data. 

Notwithstanding the paucity of DFT values of $C_{ij}\neq C_{11}$ in the extant literature, force fields that pass the previous criteria 
will be also analyzed for the agreement between the MD values of $C_{ij}$ (with $i,j\neq 11$) and the DFT data for the same $C_{ij}$ available in the literature. 

Henceforth, to differentiate between MD results and DFT calculations, the following notation will be employed: the acronyms $C_{ij}^{\mbox{\tiny{MD}}}$ and $C_{ij}^{\mbox{\tiny{DFT}}}$ are used to denote the respective concepts. 

\section{Force Field Test Results}
\label{sec3}

\indent In order to evaluate the capacity of each potential to simulate titanium MXenes, the results for the tests will be organized in two parts: 
\textit{structural properties} (the results from the first and second tests) and \textit{elastic properties} (the results from the third test). 
In each part, the results for each potential, COMB3, REAXFF and MEAM, are presented and discussed.

As previously delineated, the evaluation of force fields commences with their capacity to yield an optimized structure exhibiting the anticipated crystalline form (subsection \ref{1stcriterion}). 
Subsequently, a comparison will be made between the values of the lattice parameter, $a_0$, and thickness, $t$, of the optimized structures and the DFT values reported in the literature (see Table \ref{TSCL} and subsection \ref{2ndcriterion}).  

Finally, the {\it elastic properties} are examined, and the values of $C_{11}$, $C_{12}$, $C_{22}$, and $C_{66}$ obtained from MD simulations with force fields that satisfied the previously outlined criteria are presented and discussed in comparison to the DFT values reported in the literature (see Table~\ref{TECL} and Subsection \ref{3rdcriterion}). 
The stability and symmetry criteria, as indicated by the equations (\ref{conditionsCij}), and the condition that $C_{11}\simeq C_{22}$, respectively, will be verified.

\subsection{Structural properties}
\label{sp}

The present subsection presents the results for the two first and second tests, as previously delineated.

\subsubsection{Results of the first test}
\label{spI}

\indent Here, the force fields that failed to produce the expected crystalline structure of some or all titanium MXenes are revealed. 
Two types of failure of a force field in providing an optimized structure of MXenes have been identified.
One is the generation of a distorted optimized structure, i.e., a structure that does not exhibit the expected crystalline feature of MXenes.
The second type of failure is of a computational nature.

Three distinct distortions of the structures were observed.
One is the flattening of the structures, resulting in an optimized structure that is one-atom thick. 
The second is the amorphization of the structure.
The third distortion consists of optimized structures presenting more than one value of the lattice parameter, $a_0$. 
With the force fields and corresponding structures for which this last kind of distortion occurred, the {\it minimization} protocol was repeated with smaller values of the convergence criterion of the kinetic energy to verify if it is a matter of computational precision. 
If the converged structure remains continues to exhibit substantial differences in the values of $a_0$, the force field is  deemed inadequate for the corresponding MXene structure. 

In the event that the simulation becomes unresponsive and/or the log files contain error messages, such as ``nan'' (not a number), at any stage during the simulation of the {\it minimization} protocol, the corresponding force field will be deemed inadequate for simulating the respective MXene structure. 
A critical aspect of these evaluations is the utilization of the same structural file corresponding to each MXene structure as the input file for this test with all force fields. 
The optimized structures that passed this test ensure that when such an error occurs, it is attributed to the corresponding force field being used.
\begin{figure}[H]
\centering
\includegraphics[scale=0.60]{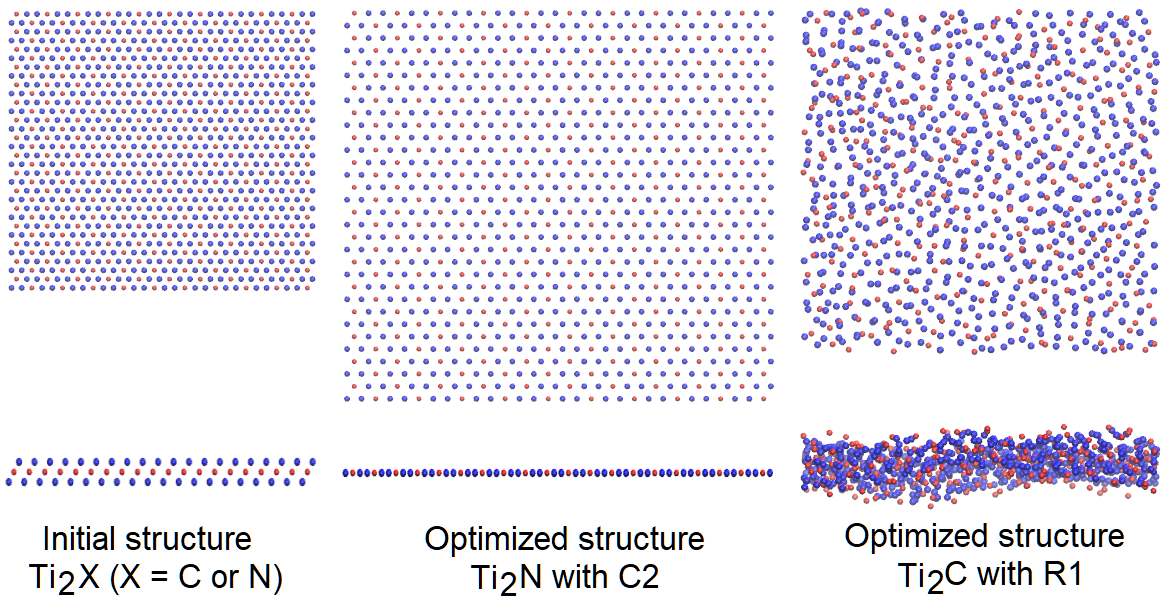}
\caption{Upper and lateral views of the initial Ti$_2$X structure (left panel) and two examples of distorted optimized structures for X = N (middle panel) and X = C (right panel). The optimized Ti$_2$N and Ti$_2$C MXene structures were obtained with C2 and R1 force fields. The initial structure is built based on lattice parameter value from the literature. Titanium atoms are shown in blue color while carbon or nitrogen atoms are drawn in red.}
\label{RCOMB3-1}
\end{figure}

Figure \ref{RCOMB3-1} shows two examples of distorted Ti$_2$X structures obtained from force fields C2 and R1.
Table \ref{tabreject} presents the types of failure that ocurred to the unsuccessful force fields. 

\begin{table}[h!]
\caption{Force fields that did not satisfy the criterion of the first test to simulate titanium MXenes, and the corresponding types of failure. See  section \ref{sec23} for the meaning of the force field abbreviations.}
\centering
\renewcommand{\arraystretch}{1.3} 
\begin{tabular}{c c c}
\hline \hline
\textbf{Structure} &  \textbf{Force field} & {\bf Kind of failure} \\
\hline \hline                    
\multirow{5}{*}{\centering Ti$_2$C} 
                               & C1  & Different values of $a_0$  \\
                               & C2 & flattening\\
                               & R1 & amorphization 
                               \\
                               & R6 & amorphization
                               \\
                               
\hline
\multirow{1}{*}{\centering Ti$_3$C$_2$} 
                               & R5 & amorphization
                               \\
                               
%
\hline \hline
\multirow{7}{*}{\centering Ti$_2$N} 
                               & C1 & Different values of $a_0$ \\
                               & C2 & Different values of $a_0$ \\
                               & R1 & nan \\
                               & R2 & nan \\
                               & R4 & nan \\
                               & R5 & nan \\
                               & M2 & nan \\
                               
\hline
\multirow{4}{*}{\centering Ti$_3$N$_2$} 
                               & R1 & nan \\
                               & R2 & nan \\
                               & R4 & nan \\
                               & R5 & nan \\
                               
\hline
\multirow{5}{*}{\centering Ti$_4$N$_3$} 
                               & C1 & Different values of $a_0$ \\
                               & R1 & nan \\
                               & R2 & nan \\
                               & R4 & nan \\
                               & R5 & nan \\
                               
\hline \hline
\end{tabular}%
\label{tabreject}
\end{table}

\subsubsection{Results of the second test}
\label{spII}

\indent The first quantitative assessment of the force fields intended to simulate titanium MXenes is founded upon a comparative analysis of the values for the lattice parameter, $a_0$, and the thickness, $t$, between those derived from DFT calculations found in the literature and those obtained from the optimized structures that passed the previous test. 

The results for each potential, COMB3, REAXFF and MEAM, are presented in subsequent subsections.  

\subsubsection{COMB3}
\label{spIIcomb3}

\indent As shown in Table \ref{tabreject}, the first test revealed that COMB3 was not effective in optimizing the thinnest titanium MXenes. 
However, COMB3 has successfully passed the initial test and has qualitatively optimized the following MXenes: Ti$_3$C$_2$, Ti$_4$C$_3$, Ti$_3$N$_2$ and Ti$_4$N$_3$. 
The structural properties, $a_0$ and $t$, of these structures obtained from the two COMB3 force-fields, C1 and C2, are shown in Table \ref{TCSC}, together with the corresponding DFT calculations found in the literature. 
Table \ref{TCSC} also shows the percentage differences $\Delta a_0/a_0^{\mbox{\tiny{DFT}}}$ and $\Delta t/t^{\mbox{\tiny{DFT}}}$, between the quantities obtained from MD and the closest extreme of the range of DFT values for the same structure. 
As previously stated in section \ref{2ndcriterion}, the criteria for identifying the optimal force fields in the second test are as follows: $\Delta a_0/a_0^{\mbox{\tiny{DFT}}}$  must be less than 5 \%, and $\Delta t/t^{\mbox{\tiny{DFT}}}$ must be less than 33.3 \%.

\begin{table}[h!]
\caption{The lattice parameter, $a_0$, and thickness, $t$, given in \AA, for the optimized titanium MXene structures obtained using COMB3 C1 and C2 force fields and from DFT calculations found in the literature. As the last presents different values for the same quantity, a range from the smallest to the largest  values is shown inside brackets [min--max]. The percentage differences, expressed as $\Delta a_0/a_0^{\mbox{\tiny{DFT}}}$ and $\Delta t/t^{\mbox{\tiny{DFT}}}$, in percentages, between the quantities obtained from MD and the closest extreme of the range of DFT values are also provided for the structural parameters. Force fields that are indicated in {\bf bold} font are those that satisfied the criteria established for the current evaluation. }
\centering
\renewcommand{\arraystretch}{1.3} 
\resizebox{\textwidth}{!}{%
\small
\begin{tabular}{c c c c c c c c}
\hline \hline
\textbf{Structure} &  \textbf{Force field} & $a_0^{\mbox{\tiny{DFT}}}$ & $a_0^{\mbox{\tiny{MD}}}$ & $\Delta a_0/a_0^{\mbox{\tiny{DFT}}}$ & $t^{\mbox{\tiny{DFT}}}$ & $t^{\mbox{\tiny{MD}}}$ & $\Delta t/t^{\mbox{\tiny{DFT}}}$ \\
\hline \hline
\multirow{2}{*}{\centering Ti$_3$C$_2$} 
                               & C1 & 
\multirow{2}{*}{\centering [2.91--3.10]} 
                               & 3.37 & 8.7 &
\multirow{2}{*}{\centering [4.60--4.69]}
                               & 5.36 & 14.3
                               \\
                               & C2 & 
                               & 3.67 & 18.4 &
                               & 3.51 & 23.7
                               \\  
\hline
\multirow{2}{*}{\centering Ti$_4$C$_3$} 
                               & C1 & 
\multirow{2}{*}{\centering [3.07-3.09]} 
                               & 3.39 & 9.7 &
\multirow{2}{*}{\centering [7.14-7.16]} 
                               & 8.14 & 13.7 
                               \\
                               & C2 & 
                               & 3.59 & 16.2 &
                               & 5.61 & 21.4 
                               \\  
\hline
\multirow{2}{*}{\centering Ti$_3$N$_2$} 
                               & {\bf C1} & 
\multirow{2}{*}{\centering [3.00-3.07]} 
                               & 2.89 & 3.7 &
\multirow{2}{*}{\centering [4.60-5.12]}
                               & 4.22 & 8.3 
                               \\
                               & {\bf C2} & 
                               & 2.89 & 3.7 &
                               & 4.22 & 8.3 
                               \\  
\hline
\multirow{1}{*}{\centering Ti$_4$N$_3$} 
                               & {\bf C2} & 
\multirow{1}{*}{\centering 2.99} 
                               & 2.86 & 4.3 &
\multirow{1}{*}{\centering 7.36}
                               & 6.47 & 12.1
                               \\  
\hline \hline
\end{tabular}%
}
\label{TCSC}
\end{table}

The MD results for the thickness values were found to meet the criterion established for the second test.  
However, the results for the lattice parameter, $a_0$, demonstrate that the COMB3 potential is only capable of simulating the structural properties of Ti$_3$N$_2$ (both C1 and C2) and Ti$_4$N$_3$ (only C2). 
For Ti$_3$C$_2$ and Ti$_4$C$_3$, COMB3 exhibits discrepancies in $a_0$ greater than 5 \%. 

\subsubsection{REAXFF}

Table \ref{TRSC} shows the values of $a_0$ and $t$ for the titanium MXenes that satisfied the criteria of the first test with the REAXFF force fields. 
\begin{table}[h!]
\caption{The lattice parameter, $a_0$, and thickness, $a_0$, and thickness, $t$, given in \AA, for the optimized titanium MXene structures obtained using REAXFF R1,R2,R3,R4,R5 and R6 force fields and from DFT calculations found in the literature. As the last presents different values for the same quantity, a range from the smallest to the largest values is shown inside brackets [min--max]. The percentage differences, expressed as $\Delta a_0/a_0^{\mbox{\tiny{DFT}}}$ and $\Delta t/t^{\mbox{\tiny{DFT}}}$, in percentages, between the quantities obtained from MD and the closest extreme of the range of DFT values are also provided for the structural parameters. Force fields that are indicated in {\bf bold} font are those that satisfied the criteria established for the current evaluation. The symbol ``--'' in $\Delta a_0/a_{0}^{\mbox{\tiny{DFT}}}$ indicates that the corresponding values of $a_0$ and/or $t$ are within the range of the corresponding DFT values.}
\centering
\renewcommand{\arraystretch}{1.3}
\resizebox{\textwidth}{!}{%
\small
\begin{tabular}{c c c c c c c c}
\hline \hline
\textbf{Structure} &  \textbf{Force field} & $a_0^{\mbox{\tiny{DFT}}}$ & $a_0^{\mbox{\tiny{MD}}}$ & $\Delta a_0/a_0^{\mbox{\tiny{DFT}}}$ & $t^{\mbox{\tiny{DFT}}}$ & $t^{\mbox{\tiny{MD}}}$ & $\Delta t/t^{\mbox{\tiny{DFT}}}$ \\
\hline \hline
\multirow{4}{*}{\centering Ti$_2$C} 
                               &  {\bf R2} & 
\multirow{4}{*}{\centering [3.00--3.04]} 
                               & 2.94 & 2.0 &
\multirow{4}{*}{\centering [2.23--2.31]}
                               & 2.33 & 0.9
                                \\                               
                               & R3 & 
                               & 2.77 & 7.7 &
                               & 6.85 & 196.5
                               \\                               
                               & {\bf R4} & 
                               & 3.03 & -- &
                               & 2.53 & 10.0
                               \\  
                               & {\bf R5} & 
                               & 2.97 & 1.0 &
                               & 3.08 & 33.3
                               \\  
\hline
\multirow{5}{*}{\centering Ti$_3$C$_2$} 
                               &  {\bf R1} & 
\multirow{5}{*}{\centering [2.91--3.10]} 
                               & 3.18 & 2.6 &
\multirow{5}{*}{\centering [4.60--4.69]}
                               & 4.97 & 6.0
                               \\
                               &  {\bf R2} & 
                               & 3.04 & -- &
                               & 5.04 & 7.5
                               \\  
                               & R3 & 
                               & 2.77 & 4.8 &
                               & 12.92 & 175.4
                               \\  
                               & {\bf R4} & 
                               & 3.04 & -- &
                               & 5.05 & 7.7
                               \\ 
                               &  {\bf R6} & 
                               & 3.15 & 1.6 &
                               & 5.15 & 9.8
                               \\ 
\hline
\multirow{6}{*}{\centering Ti$_4$C$_3$} 
                               & {\bf R1} & 
\multirow{6}{*}{\centering [3.07--3.09]} 
                               & 3.16 & 2.3 &
\multirow{6}{*}{\centering [7.14--7.16]}
                               & 7.61 & 6.3
                               \\
                               & {\bf R2} & 
                               & 3.04 & -- &
                               & 7.59 & 6.0
                               \\  
                               & R3 & 
                               & 2.78 & 9.4 &
                               & 18.97 & 165.0
                               \\  
                               & {\bf R4} & 
                               & 3.05 & -- &
                               & 7.55 & 5.4
                               \\ 
                               & R5 & 
                               & 2.94 & 4.2 &
                               & 10.69 & 49.3
                               \\
                               & {\bf R6} & 
                               & 3.13 & 1.3 &
                               & 8.97 & 25.3
                               \\  
\hline
\multirow{1}{*}{\centering Ti$_2$N} 
                               & {\bf R3} & 
\multirow{1}{*}{\centering [2.98--3.03]} 
                               & 2.99 & -- &
\multirow{1}{*}{\centering [2.27--2.31]}
                               & 2.12 & 6.6
                               \\
\hline
\multirow{1}{*}{\centering Ti$_3$N$_2$} 
                               & {\bf R3} & 
\multirow{1}{*}{\centering [3.00--3.07]} 
                               & 3.01 & -- &
\multirow{1}{*}{\centering [4.60--5.12]}
                               & 4.64 & --
                               \\
\hline
\multirow{1}{*}{\centering Ti$_4$N$_3$} 
                               & {\bf R3} & 
\multirow{1}{*}{\centering 2.99} 
                               & 3.01 & 0.7 &
\multirow{1}{*}{\centering 7.36}
                               & 7.14 & 3.0
                               \\
\hline \hline
\end{tabular}%
}
\label{TRSC}
\end{table}
As with the COMB3 force fields, a single REAXFF force field is not capable of simulating all titanium MXenes. 
In contrast to the findings observed with COMB3, certain REAXFF force fields yielded unsatisfactory values for the thickness, $t$, of some MXenes. 
Structures that were optimized with R3 for all titanium carbide MXenes and with R5 for Ti$_4$C$_3$ exhibited remarkably large values of thickness.

In contrast, several REAXFF force fields are shown to satisfy the criteria for $a_0$ and $t$ for various titanium MXenes. 
The following REAXFF force fields have been demonstrated to accurately simulate the structural properties of all three titanium carbide (all three titanium nitride) MXenes: R2 and R4 (R3).

The R1 and R6 REAXFF force fields did not successfully pass the first test to simulate Ti$_2$C and all three nitride MXenes; however, they were demonstrated to accurately simulate the structures of Ti$_3$C$_2$ and Ti$_4$C$_3$. 
The R5 is shown to meet the current criteria for simulating only the structure of Ti$_2$C. 
The R3 force field has been demonstrated to be inadequate for simulating all titanium carbide MXenes; however, it has been very effective for simulating all nitride MXenes.

In the context of MXenes whose structural characteristics can be accurately reproduced by REAXFF force fields,  Ti$_2$C has been shown to be correctly simulated by R2, R4 and R5. 
It has been demonstrated that Ti$_3$C$_2$ and Ti$_4$C$_3$ can be simulated by R1, R2, R4 and R6 force fields. 
The structural integrity of all three nitride MXenes has been demonstrated to be accurately simulated by R3.

With respect to the thickness of the optimized structures, R1, R2, and R4 yielded the most favorable values in comparison to the DFT data for Ti$_3$C$_2$, Ti$_2$C and Ti$_4$C$_3$, respectively. 
It is noteworthy that R3 furnished precise thickness values for nitride MXenes.

\subsubsection{MEAM}

Table \ref{TMSC} provides the results for the structural parameters, $a_0$ and $t$, from the MD simulations with the MEAM potential. 
In contrast to the preceding potentials, the MEAM force fields demonstrate the capacity to simulate the structural properties of all 
titanium MXenes with good precision, except for the Ti$_2$N structure. 
The agreement between the MD and DFT values of the lattice parameter, $a_0$, is remarkable for all MXenes that passed the first test. 
However, this correlation is not observed for the thickness of all structures. 
While the thickness values of Ti$_3$N$_2$ and Ti$_4$N$_3$ obtained with the M2 force field are in close agreement with the DFT data, for the titanium carbide MXenes, the thickness values of the optimized structures with the M1 force field are within the limit of the proposed criterion of 33.3 \% of the maximum percentage difference. 
It is noteworthy that the parameterization of M2 to simulate bulk TiN is capable of providing reliable values of $t$ for Ti$_3$N$_2$ and Ti$_4$N$_3$.  

The optimized Ti$_2$C structure obtained with the M1 force field exhibited two close values of lattice parameter: $a_0=3.01$ and 3.02. 
Running the {\it minimization} protocol with a reduced convergence criterion value was insufficient to obtain an optimized structure with only one value $a_0$. 
Given the negligible discrepancy between the two values (approximately 0.33 \%), it was posited that the M1 force field would be adequate for simulating the structure of Ti$_2$C. 
The mean value of these parameters was calculated to determine the MD value for  $a_0$ of the Ti$_2$C structure from M1 force field. 

\begin{table}[h!]
\caption{The lattice parameter, $a_0$, and thickness, $t$, given in \AA, for the optimized titanium MXene structures obtained using MEAM M1 and M2 force fields and from DFT calculations found in the literature. As the last presents different values for the same quantity, a range from the smallest to the largest values is shown inside brackets [min--max]. The percentage differences, expressed as $\Delta a_0/a_0^{\mbox{\tiny{DFT}}}$ and $\Delta t/t^{\mbox{\tiny{DFT}}}$, in percentages, between the quantities obtained from MD and the closest extreme of the range of DFT values are also provided for the structural parameters. Force fields that are indicated in {\bf bold} font are those that satisfied the criteria established for the current evaluation. The symbol ``--'' in $\Delta a_0/a_{0}^{\mbox{\tiny{DFT}}}$ indicates that the corresponding values of $a_0$ and/or $t$ are within the range of the corresponding DFT values.}
\vspace{-7mm}
\begin{center}
\renewcommand{\arraystretch}{1.3} 
\resizebox{\textwidth}{!}
{
\small
\begin{tabular}{c c c c c c c c}
\hline \hline
\textbf{Structure} & \textbf{Force field} & $a_0^{\mbox{\tiny{DFT}}}$ & $a_0^{\mbox{\tiny{MD}}}$ & 
$\Delta a_0/a_0^{\mbox{\tiny{DFT}}}$ & $t^{\mbox{\tiny{DFT}}}$ & $t^{\mbox{\tiny{MD}}}$ & 
$\Delta t/t^{\mbox{\tiny{DFT}}}$ \\
\hline \hline
\multirow{1}{*}{\centering Ti$_2$C} 
                               & M1 & 
\multirow{1}{*}{\centering [3.00--3.04]} 
                               & 
3.02$^{*}$ 
                               & -- &
\multirow{1}{*}{\centering [2.23 - 2.31]}
                               & 3.03 & 31.2
                               \\  
\hline
\multirow{1}{*}{\centering Ti$_3$C$_2$} 
                               & M1 & 
\multirow{1}{*}{\centering [2.91--3.10]} 
                               & 3.01 & -- &
\multirow{1}{*}{\centering [4.60--4.69]} 
                               & 5.84 & 24.6 
                               \\
\hline                              
\multirow{1}{*}{\centering Ti$_4$C$_3$} 
                               & M1 & 
\multirow{1}{*}{\centering [3.07--3.09]} 
                               & 3.03 & 1.3 &
\multirow{1}{*}{\centering [7.14--7.16]} 
                               & 8.46 & 18.2 
                               \\
\hline                              
\multirow{1}{*}{\centering Ti$_3$N$_2$} 
                               & M2 & 
\multirow{1}{*}{\centering [3.00--3.07]} 
                               & 3.02 & -- &
\multirow{1}{*}{\centering [4.60--5.12]}
                               & 4.56 & 0.9
                               \\  
\hline                              
\multirow{1}{*}{\centering Ti$_4$N$_3$} 
                               & M2 & 
\multirow{1}{*}{\centering 2.99} 
                               & 3.02 & 1.0 &
\multirow{1}{*}{\centering 7.36}
                               & 6.99 & 5.0
                               \\   
\hline \hline
\end{tabular}
}
\end{center}
\vspace{-4mm}
{\footnotesize 
\begin{spacing}{1.0}
\noindent * This value is an average between 3.01 and 3.02 that were obtained from the {\it minimization} protocol, even using smaller convergence criterion. As these two values are very close one to each other, we considered this force field and the $a_0$ value given by the average.
\end{spacing}
}
\label{TMSC}
\end{table}

\subsubsection{Conclusion remarks}

\begin{figure}[H]
\centering
\includegraphics[scale=0.21]{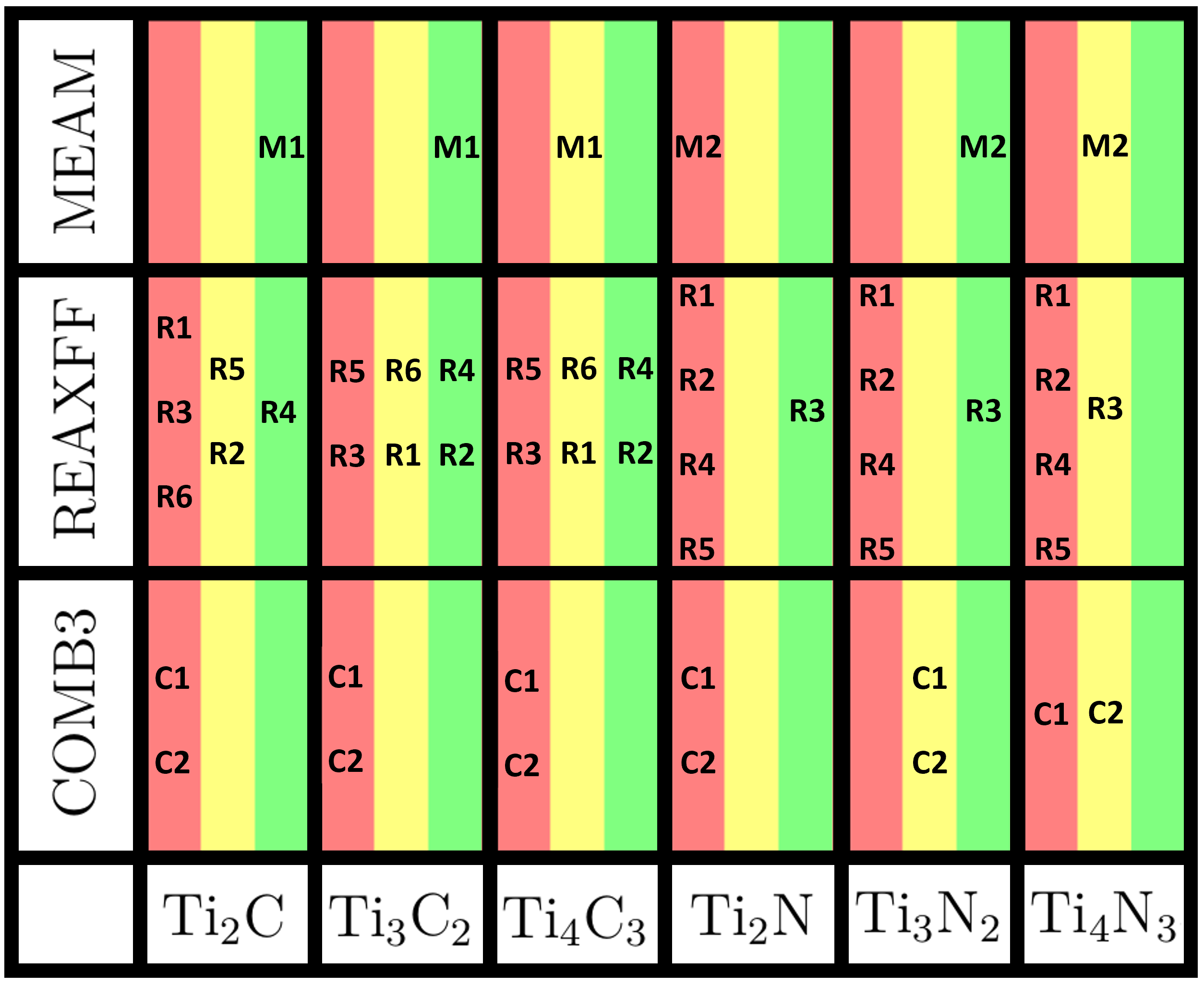}
\caption{Summary of the main MD results from the {\it structural properties} section \ref{spI}. Force fields written inside green, orange and red columns are: those whose values of lattice parameter, $a_0$, fall within the range of DFT calculations for the corresponding MXene structure; those whose values of $a_0$ were no more than 5 \% the minimum or maximum of DFT calculations; and those which did not satisfy the structural criteria.}
\label{fig4}
\end{figure}

Figure \ref{fig4} was prepared to summarize the results of the {\it structural properties} tests. 
The figure illustrates a table showing the force fields inside colored columns coded according to their ability to simulate the structural properties of corresponding titanium MXenes. 
The force fields located in the green columns are those that provided values of lattice parameters, $a_0$, within the range of DFT calculations found in the literature for the corresponding optimized structures. 
The force fields located in the yellow columns represent those that provided optimized structures with lattice parameters, $a_0$, falling within 5 \% of the minimum and maximum values of DFT data reported in the literature, for the corresponding optimized structure. 
The force fields located in the red columns are those that failed the {\it structural properties} tests. 

\subsection{Elastic Constants}
\label{sec32}

In this section, the elastic constants of the MXenes obtained from the force fields that passed the previous tests are presented and discussed. The elastic constants were obtained using the {\it stretching} protocol described in section \ref{sec231}.
The accuracy of the results will be analyzed in terms of the available DFT calculations for the elastic constants of the titanium MXenes, as shown in Table \ref{TECL}.

As outlined in the preceding section, the results are presented for each of the potentials, COMB3, REAXFF, and MEAM.

\subsubsection{COMB3}
\label{sec321}

The COMB3 force fields that passed the previous tests are those that optimized the structures of Ti$_3$N$_2$ (both C1 and C2) and Ti$_4$N$_3$ (only C2) MXenes. 
Table \ref{tab:COMB3C11C12} shows the elastic constants of these structures as obtained from the MD simulations with C1 and C2 force fields, $C_{ij}^{\mbox{\tiny{MD}}}$, and from the DFT data available in the literature, $C_{ij}^{\mbox{\tiny{DFT}}}$. 

\begin{table}[h!]
\caption{Elastic constants, $C_{11}$, $C_{12}$, $C_{22}$ and $C_{66}$, in N/m, of the optimized titanium MXene structures obtained with the COMB3 force fields that passed the previous tests, and comparison with DFT data. $C_{ij}^{\mbox{\tiny{MD}}}$ and $C_{ij}^{\mbox{\tiny{DFT}}}$ are the values from the MD simulations and DFT calculations found in the literature, respectively. }
\centering
\renewcommand{\arraystretch}{1.3}
\resizebox{\textwidth}{!}{%
\small
\begin{tabular}{c c c c c c c c c c}
\hline \hline
\textbf{Structure} & \textbf{Force field} & $C_{11}^{\mbox{\tiny{DFT}}}$ & $C_{11}^{\mbox{\tiny{MD}}}$ & $C_{12}^{\mbox{\tiny{DFT}}}$ & $C_{12}^{\mbox{\tiny{MD}}}$ & $C_{22}^{\mbox{\tiny{DFT}}}$ & $C_{22}^{\mbox{\tiny{MD}}}$ & $C_{66}^{\mbox{\tiny{DFT}}}$ & $C_{66}^{\mbox{\tiny{MD}}}$\\
\hline \hline
\multirow{2}{*}{Ti$_3$N$_2$} 
                             & C1 & 
\multirow{2}{*}{263}                       
                             & 635 & 
\multirow{2}{*}{--}                            
                             & 133 &
\multirow{2}{*}{--}                            
                             & 637 &
\multirow{2}{*}{--}                            
                             & 251 \\
                             & C2 & 
                             & 635 & 
                             & 134 &
                             & 638 &
                             & 251 \\
\hline
\multirow{1}{*}{Ti$_4$N$_3$}
                             & C2 & 
\multirow{1}{*}{369}                       
                             & 965 & 
\multirow{1}{*}{--}                            
                             & 196 &
\multirow{1}{*}{--}                            
                             & 964 &
\multirow{1}{*}{--}                            
                             & 384 \\
\hline \hline
\end{tabular}%
}
\label{tab:COMB3C11C12}
\end{table}

It is evident that for all structures,  $C_{11}^{\mbox{\tiny{MD}}}\simeq C_{22}^{\mbox{\tiny{MD}}}$ (the largest discrepancy is less than 0.5 \%), and the conditions stipulated by equation (\ref{conditionsCij}) are met. 
However, a notable discrepancy exists between the MD and DFT results. 
Despite its capacity to simulate the structures of Ti$_3$N$_2$ and Ti$_4$N$_3$, COMB3 overestimates their rigidity. 

\begin{table}[h!]
\caption{Elastic constants, $C_{11}$, $C_{12}$, $C_{22}$ and $C_{66}$ in N/m, of the optimized titanium MXene structures obtained with the REAXFF force fields that passed the previous tests, and comparison with DFT data. $C_{ij}^{\mbox{\tiny{MD}}}$ and $C_{ij}^{\mbox{\tiny{DFT}}}$ are the values from the MD simulations and DFT calculations found in the literature, respectively. }
\centering
\renewcommand{\arraystretch}{1.3}
\resizebox{\textwidth}{!}{%
\small
\begin{tabular}{c c c c c c c c c c}
\hline \hline
\textbf{Structure} & \textbf{Force field} & $C_{11}^{\mbox{\tiny{DFT}}}$ & $C_{11}^{\mbox{\tiny{MD}}}$ & $C_{12}^{\mbox{\tiny{DFT}}}$ & $C_{12}^{\mbox{\tiny{MD}}}$ & $C_{22}^{\mbox{\tiny{DFT}}}$ & $C_{22}^{\mbox{\tiny{MD}}}$ & $C_{66}^{\mbox{\tiny{DFT}}}$ & $C_{66}^{\mbox{\tiny{MD}}}$ \\   
\hline \hline                                            
\multirow{3}{*}{Ti$_2$C} 
                             &  R2 & 
\multirow{3}{*}{[130 - 151]} 
                             &  236   &  
\multirow{3}{*}{[32 - 37]} 
                             &  -91   & 
\multirow{3}{*}{153}                             
                             &  228   &
\multirow{3}{*}{58} 
                             & 168   \\                         
                             &  R4   &                   
                             &  264   & 
                             &  74   &                  
                             &  262   & 
                             &  95   \\
                             &  R5   &                   
                             &  47   & 
                             &  1   &                  
                             &  45   & 
                             &  26   \\
\hline
\multirow{4}{*}{Ti$_3$C$_2$} 
                             & R1 & 
\multirow{4}{*}{[219 - 253]} 
                             & 143 &   
\multirow{4}{*}{40} 
                             & 100 & 
\multirow{4}{*}{257}                              
                             &  142   &
\multirow{4}{*}{107}                             
                             &  21   \\
                             & R2 &                   
                             & 391    & 
                             &  10   &                  
                             &  390   & 
                             &  190   \\
                             & R4 &                   
                             &  391   & 
                             &  105   &                  
                             &  388   & 
                             &  142   \\
                             & R6  &                   
                             & 120    & 
                             & 80    &                  
                             &  119   & 
                             &  20   \\
\hline
\multirow{4}{*}{Ti$_4$C$_3$} 
                             & R1 & 
\multirow{4}{*}{[312 - 366]} 
                             & 182 & 
\multirow{4}{*}{49} 
                            &  103    &
\multirow{4}{*}{306} 
                            &  181    &
\multirow{4}{*}{130}                             
                            &  39    \\
                            &  R2   &                   
                            &  531   & 
                            &  -0.1   &                  
                            &  530   & 
                            &  266   \\                   
                            & R4 & 
                            &  521   & 
                            &  136   &                  
                            &  517   & 
                            &  191   \\
                            & R6 &                   
                            &  158   & 
                            &   96  &                  
                            &  157   & 
                            &  36   \\
\hline
\multirow{1}{*}{Ti$_2$N} 
                             & R3 & 
\multirow{1}{*}{[150 - 154]} 
                             & 189 &  
\multirow{1}{*}{41} 
                             &  91 & 
\multirow{1}{*}{--} 
                             &   188   &
\multirow{1}{*}{--} 
                             &   49   \\                                               
\hline
\multirow{1}{*}{Ti$_3$N$_2$} 
                             & R3 & 
\multirow{1}{*}{263} 
                             & 418 &
\multirow{1}{*}{--}                              
                             &   161   &
\multirow{1}{*}{--}                              
                             &   417   &
\multirow{1}{*}{--}                              
                             &   128   \\
\hline
\multirow{1}{*}{Ti$_4$N$_3$} 
                             &   R3 & 
\multirow{1}{*}{369} 
                             &  578  & 
\multirow{1}{*}{--} 
                             &  206   & 
\multirow{1}{*}{--}                              
                             &  573   &
\multirow{1}{*}{--}                              
                             &  185   \\                          
\hline \hline
\end{tabular}%
}
\label{tab:REAXC11C12}
\end{table}

\subsubsection{REAXFF}
\label{sec322}

The $C_{ij}$ results for the REAXFF force fields are presented in Table \ref{tab:REAXC11C12}. 
It can be seen that all force fields satisfied the symmetry criterion, $C_{22}^{\mbox{\tiny{MD}}}\simeq C_{11}^{\mbox{\tiny{MD}}}$ (the largest discrepancy is less than 5 \%) and the stability criterion given by equation (\ref{conditionsCij}).  

Table \ref{tab:REAXC11C12} shows a substantial variation in the $C_{ij}^{\mbox{\tiny{MD}}}$ values among the different force fields. 
For Ti$_2$C, for instance, the discrepancy between the outcomes of $C_{11}^{\mbox{\tiny{MD}}}$ from R4 and R5 force fields is as large as 217 N/m, between R4 and R5 results, a value that exceeds the absolute $C_{11}^{\mbox{\tiny{DFT}}}$ ones. 
For the $C_{12}^{\mbox{\tiny{MD}}}$, discrepancies are observed not only in the absolute values but also in the sign.
 
Among the REAXFF force fields, R4 provides the closest values between $C_{ij}^{\mbox{\tiny{MD}}}$ and $C_{ij}^{\mbox{\tiny{DFT}}}$ for Ti$_2$C, Ti$_3$C$_2$ and Ti$_4$C$_3$ structures. 
R3 is the sole REAXFF force field capable of simulating the structure and mechanical properties of all three titanium nitride MXenes. 
Indeed, R3 is the sole force field in the present study that possesses the capacity to simulate Ti$_2$N MXenes. 
For this structure, R3 provides a reasonable value for the $C_{11}^{\mbox{\tiny{MD}}}$, exhibiting a 22.7 \% of discrepancy with the $C_{11}^{\mbox{\tiny{DFT}}}$. 
However, a significant discrepancy is observed between $C_{12}^{\mbox{\tiny{MD}}}$ and $C_{12}^{\mbox{\tiny{DFT}}}$.

\subsubsection{MEAM}
\label{sec323}

The elastic constants, $C_{ij}^{\mbox{\tiny{MD}}}$, obtained through the use of MEAM force fields are presented in Table \ref{tab:MEAMC11C12}. 
Their values satisfy the stability criterion. 
A comparison of $C_{11}^{\mbox{\tiny{MD}}}$ and $C_{11}^{\mbox{\tiny{DFT}}}$ reveals a high degree of agreement between them for the M1 force field.  
However, the symmetry test fails for the Ti$_2$C structure.
The percentage difference between $C_{11}^{\mbox{\tiny{MD}}}$ and $C_{22}^{\mbox{\tiny{MD}}}$ is greater than 20 \%. 
This outcome may be associated with the force field limitation observed in the analysis of the second test, wherein M1 did not provide a unique value for the lattice parameter, $a_0$. 
For Ti$_3$C$_2$ and Ti$_4$C$_3$ structures, the discrepancy between $C_{11}^{\mbox{\tiny{MD}}}$ and $C_{22}^{\mbox{\tiny{MD}}}$ is found to be significantly reduced. 
Given the relatively low differences between $C_{11}^{\mbox{\tiny{MD}}}$ and $C_{11}^{\mbox{\tiny{DFT}}}$ for Ti$_3$C$_2$ and Ti$_4$C$_3$ structures,  M1 is regarded as an adequate force field for simulating their structure and elastic properties. 

For Ti$_3$N$_2$ and Ti$_4$N$_3$ structures, the symmetry and stability criteria are both satisfied. 
Due to the small differences between $C_{11}^{\mbox{\tiny{MD}}}$ and $C_{11}^{\mbox{\tiny{DFT}}}$ for these structures, M2 is regarded as the most suitable force field for simulating them.

Therefore, it can be concluded that the MEAM potential adequately describes the structure and elastic behavior of the Ti$_3$X$_2$ and Ti$_4$X$_3$ (X = C,N) MXene structures.

\begin{table}[h!]
\caption{Elastic constants, $C_{11}$, $C_{12}$, $C_{22}$ and $C_{66}$ in N/m, of the optimized titanium MXene structures obtained with the MEAM force fields that passed the previous tests, and comparison with DFT data. $C_{ij}^{\mbox{\tiny{MD}}}$ and $C_{ij}^{\mbox{\tiny{DFT}}}$ are the values from the MD simulations and DFT calculations found in the literature, respectively.}
\centering
\renewcommand{\arraystretch}{1.3}
\resizebox{\textwidth}{!}{%
\small
\begin{tabular}{c c c c c c c c c c}
\hline \hline
\textbf{Structure} & \textbf{Force field} & $C_{11}^{\mbox{\tiny{DFT}}}$ & $C_{11}^{\mbox{\tiny{MD}}}$ & $C_{12}^{\mbox{\tiny{DFT}}}$ & $C_{12}^{\mbox{\tiny{MD}}}$ & $C_{22}^{\mbox{\tiny{DFT}}}$ & $C_{22}^{\mbox{\tiny{MD}}}$ & $C_{66}^{\mbox{\tiny{DFT}}}$ & $C_{66}^{\mbox{\tiny{MD}}}$ \\
\hline \hline
\multirow{1}{*}{Ti$_2$C} 
                             & M1 & 
\multirow{1}{*}{[130 - 151]} 
                             & 162 &  
\multirow{1}{*}{[32 - 37]} 
                             & 79 & 
\multirow{1}{*}{153} 
                             & 133 & 
\multirow{1}{*}{58} 
                             &  34  \\                            
\hline
\multirow{1}{*}{Ti$_3$C$_2$} 
                             & M1 & 
\multirow{1}{*}{[219 - 253]} 
                             & 274 &  
\multirow{1}{*}{40} 
                             & 67 & 
\multirow{1}{*}{257} 
                             & 283 & 
\multirow{1}{*}{107} 
                             & 105 \\                            
\hline
\multirow{1}{*}{Ti$_4$C$_3$} 
                             & M1 & 
\multirow{1}{*}{[312 - 366]} 
                             & 365 & 
\multirow{1}{*}{49} 
                             & 80 & 
\multirow{1}{*}{306} 
                             & 368 & 
\multirow{1}{*}{130} 
                             & 143 \\                            
\hline
\multirow{1}{*}{Ti$_3$N$_2$} 
                             & M2 & 
\multirow{1}{*}{263} 
                             & 193 & 
\multirow{1}{*}{--} 
                             & 35 &  
\multirow{1}{*}{--} 
                             & 192 & 
\multirow{1}{*}{--} 
                             & 79  \\                             
\hline
\multirow{1}{*}{Ti$_4$N$_3$} 
                             & M2 & 
\multirow{1}{*}{369} 
                             & 303 &  
\multirow{1}{*}{--} 
                             & 57 & 
\multirow{1}{*}{--} 
                             & 302 &  
\multirow{1}{*}{--} 
                             & 123  \\                          
\hline \hline
\end{tabular}%
}
\label{tab:MEAMC11C12}
\end{table}

\section{Elastic Properties of Titanium MXenes}
\label{sec4}

\indent The force fields that passed the three previous tests and are thus considered to correctly simulate structural results and reasonable elastic constants for each titanium MXene are summarized below:
 
\begin{enumerate}
\item[Ti$_2$C ] -- R4;
\item[Ti$_3$C$_2$] -- R4 and M1;
\item[Ti$_4$C$_3$] -- R4 and M1;
\item[Ti$_2$N ] -- R3;
\item[Ti$_3$N$_2$] -- M2;
\item[Ti$_4$N$_3$] -- M2.
\end{enumerate}

In this section, the results for the Young's modulus, linear compressibility, Poisson's ratio, and shear modulus of all titanium MXene structures examined in this study are presented. 
Table \ref{TEMQ} presents the values of these elastic quantities for each structure. 
The values were calculated using equation (\ref{2}) and the $C_{ij}$ obtained from MD simulations with the effective force fields. 
Furthermore, Table \ref{TEMQ} presents the results obtained from DFT calculations found in the literature.

\begin{table}[h!]
\caption{Elastic quantities, $E_x$, $E_y$, $\beta_x$, $\beta_y$, $\nu_{xy}$, $\nu_{yx}$ and $G$, of MXenes for the force fields that satisfied the previous criteria for the structural properties and elastic constants. The values in the literature, obtained from DFT calculations, either given directly or calculated from the elastic constants using equation (\ref{2}) are also shown for each reference. $E_x$, $E_y$ and $G$ are given in $N/m$, while $\beta_x$ and $\beta_y$ in $10^{-3}(N/m)^{-1}$. 
}
\vspace{-7mm}
\begin{center}
\renewcommand{\arraystretch}{0.35}
\setlength{\tabcolsep}{12pt}
\small
\begin{tabular}{c c c c c c c c c}
\hline \hline \\
\textbf{Structure}  & \textbf{Source} & $E_x$ & $E_y$ & $\beta_x$ & $\beta_y$ & $\nu_{xy}$ & $\nu_{yx}$ & $G$  \\ \\
\hline \hline
\\
\multirow{16}{*}{\centering Ti$_2$C} 
                    & R4: & 243 & 241 & 2.9  & 3.0 & 0.282 & 0.280 & 95 \\ \\
                   \cline{2-9}               \\
            & Ref.~\cite{DFTE2}: & 130
            &  & 6.3$^\alpha$ &   & 0.23 &  & 50$^\alpha$ \\
            & Ref.~\cite{DFTM2}:       &  130 &   &    5.9$^\beta$    &       &  0.23$^\beta$ &    & 52$^\beta$ \\
            & Ref.~\cite{DFTM6}:        & 
            139$^a$  &  143$^a$  &    &        &   &      \\
            & Ref.~\cite{DFTM9}:     &  125 &   &    5.8$^\beta$    &       &  0.27$^\beta$ &    & 49$^\beta$ \\
            & Ref.~\cite{DFTE6}: & 119 
            &  & 5.7$^\alpha$ &  & 0.266 &   &  51$^\alpha$ \\    
            & Ref.~\cite{DFTM3}: & 143  &  144  &    5.4$^\beta$    &  5.3$^\beta$      & 0.231  &  0.233  &  54 \\
\\                               
\hline \hline 
\\
\multirow{15}{*}{\centering  Ti$_3$C$_2$} 
                & R4: & 363 & 360 & 2.0  & 2.0 & 0.269 & 0.271 & 144
                \\  \\
                & M1:  & 258 & 267 & 3.0  & 2.8 & 0.245 & 0.237 &  106 
                \\  \\   
                \cline{2-9}
                \\    
        & Ref.~\cite{DFTE2}: & 231
        &  & 3.7$^\alpha$ &  &  0.17 &  & 96$^\alpha$ \\
        & Ref.~\cite{PRB2016}: & 228 
        & 227 & 3.4$^\alpha$ & 3.4$^\alpha$ & 0.227 & 0.266 & 103 \\
        & Ref.~\cite{DFTM6}: &  248$^b$ & 264$^b$ &   &  &  &   &   \\
        & Ref.~\cite{DFTE6}: & 207
        &  &  3,7$^\alpha$ &   &  0.241 &   & 83$^\alpha$  \\
        & Ref.~\cite{DFTM3}: & 
        247  & 251 & 3.4$^\beta$ & 3.4$^\beta$ &  0.154  &  0.156 &  105 \\
\\
\hline \hline
\\
\multirow{13}{*}{\centering  Ti$_4$C$_3$} 
                & R4: & 485 & 481 & 1.5  & 1.5 & 0.261 & 0.263 & 194  \\
                                \\
                & M1: & 348 & 350 & 2.2 & 2.2 & 0.219 & 0.217 & 143  
                \\ \\
                \cline{2-9}
                \\ 
    & Ref.~\cite{DFTM6}: &  355$^{c}$ & 373$^{c}$ &  &  &   &  &  \\
    & Ref.~\cite{DFTE6}: & 308
    &  & 2.4$^\alpha$ &  & 0.250 &  & 123$^\alpha$ \\
    & Ref.~\cite{DFTM3}: &  305  & 298 & 2.8$^\beta$  &   2.8$^\beta$  &  0.159  &  0.155   &  132 \\
\\
\hline \hline
\\
\multirow{7}{*}{\centering Ti$_2$N} 
        & R3: & 144 & 144 & 3.6 & 3.6 & 0.481 & 0.484 & 49  \\ \\
        \cline{2-9}
        \\
        & Ref.~\cite{DFTM9}: &  143 &  & 5.1$^\beta$ & & 0.27$^\beta$ &  & 56$^\beta$ \\
        & Ref.~\cite{DFTE6}: & 123 &
        & 5.3$^\alpha$ &  & 0.271 &  & 55$^\alpha$ \\
\\                             
\hline \hline
\\
\multirow{5}{*}{\centering Ti$_3$N$_2$} 
        & M2: & 186 & 185 & 4.4  & 4.4 & 0.181 & 0.182 &  79 \\
        \\
        \cline{2-9}
        \\
        & Ref.~\cite{DFTE6}: & 238 & 
        & 2.3$^\alpha$ &  &  0.265 &   &  97$^\alpha$ \\
\\
\hline \hline
\\
\multirow{5}{*}{\centering Ti$_4$N$_3$} 
        & M2:  & 293 & 291 & 2.8  & 2.8 & 0.188 & 0.189 & 123  \\ \\
        \cline{2-9}   \\ 
        & Ref.~\cite{DFTE6}: & 347 &
        & 2.2$^\alpha$ &  & 0.250 &   & 138$^\alpha$ \\
\\                               
\hline \hline
\end{tabular}
\vspace{-4mm}
\end{center}
{\footnotesize 
\begin{spacing}{1.0}
\noindent $^a$ Original results in GPa were converted to N/m by multiplying by the average thickness of 0.23 nm.

\noindent $^b$ Original results in GPa were converted to N/m by multiplying by the average thickness of 0.46 nm.

\noindent $^c$ Original results in GPa were converted to N/m by multiplying by the average thickness of 0.71 nm.

\noindent $^\alpha$ These values are calculated by first using equations (\ref{2}) to obtain the corresponding $C_{ij}$ values, and then using the same equations to obtain the unstated elastic quantities.

\noindent $^\beta$ These quantities are calculated from the $C_{ij}$ values present in the corresponding references, using the equations (\ref{2}).
\end{spacing}
}
\label{TEMQ}
\end{table}

The elastic quantities are presented for two directions in order to verify if the method and force fields provide results coherent with the expectation from the symmetry of the titanium MXenes. 
Table \ref{TEMQ} shows that the discrepancy between quantities in the x- and y-directions, calculated from force field results, is small and comparable to those obtained from DFT calculations that did not assume the MXenes’ symmetry a priori (e.g., Ref.~\cite{DFTM3}). 

The Young's and shear moduli of all titanium carbide MXenes obtained from the REAXFF R4 and MEAM M1 force fields are higher than those from DFT calculations. 
For Ti$_3$C$_2$ and Ti$_4$C$_3$ structures, the discrepancy between MD/M1 and DFT results is less pronounced than that observed between MD/R4 and DFT.  
Consequently, titanium carbide MXenes simulated with the REAXFF force field are expected to exhibit increased rigidity.

The MD results for titanium nitride MXenes demonstrate a much better agreement with the DFT predictions for the elastic moduli, employing the REAXFF R3 (for Ti$_2$N structure) and MEAM M2 (for Ti$_3$N$_2$ and Ti$_4$N$_3$ structures) force fields. 
These two force fields possess the capacity to more accurately simulate the structure and elastic moduli of titanium nitride MXenes.

Titanium MXenes exhibit Young’s moduli comparable to other 2D materials, including graphene (350 N/m)~\cite{Cao2020NatComm}, boron-nitride (289 N/m)~\cite{Falim2017NatComm} and MoS$_2$ (126.5 N/m)\cite{Xue2024ResPhys}. 

With the exception of the Ti$_2$N structure, the values of the Poisson's ratio obtained from MD and DFT calculations demonstrate a high degree of agreement for the titanium MXenes.  
The Poisson's ratio of all titanium MXenes falls within the values for graphene (between 0.16 and 0.45)~\cite{Politano2015NanoRes} and hexagonal boron-nitride (between 0.213 and 0.413)~\cite{Siby2026JPhys}. 

An important yet unexplored parameter in the field pertains to the linear compressibility, $\beta$, of titanium MXenes.  
This parameter is indicative of the extent to which a material undergoes a change in length in response to hydrostatic pressure. 
Table \ref{TEMQ} presents the values of $\beta$ of the MXenes studied in this research. 
Given the direction-dependence of $\beta$, it is also expected that this quantity is equal along the x- and y-directions in Titanium MXenes due to symmetry reasons. 
In this regard, it is evident that all results presented in Table \ref{TEMQ} comply with the symmetry criterion.

For Ti$_2$C, Ti$_2$N and Ti$_3$C$_2$ structures, the MD values of the linear compressibility are found to be less than the DFT values, with the last structure presenting the smallest discrepancy. 

The linear compressibility values of titanium MXenes are commensurate with those reported in the extant literature.
Graphite and $\gamma$-graphyne, for  instance, exhibit linear compressibilities of approximately $\beta\simeq2.4\times10^{-3}$ (N / m)$^{-1}$~\cite{Hanf1989PRB} and $4.0\times10^{-3}$ (N / m)$^{-1}$~\cite{Kanegae2022CarbonTrends}, respectively.  
The linear compressibility of a monolayer of MoS$_2$, calculated from its $C_{ij}$ parameters~\cite{Xue2024ResPhys}, is larger, and has a value of $6.2\times10^{-3}$ (N / m)$^{-1}$.

\section{Conclusion}
\label{sec5}

The present study evaluated the ability of three MD potentials -- COMB3, REAXFF, and MEAM -- to simulate the structure and elastic properties of titanium MXenes. 
A total of ten force fields were examined; two of these were from COMB3, six were from REAXFF, and two were from MEAM. 
However, only four of these force fields adequately simulated the structural and elastic properties of titanium MXenes. 
Moreover, the efficacy of these force fields in simulating the behavior of MXenes was only partial. 
It was determined that no force field was capable of accurately simulating all titanium MXenes. 
The following force fields were found to be effective: REAXFF R4~\cite{R4} (for Ti$_2$C, Ti$_3$C$_2$ and Ti$_4$C$_3$ structures), MEAM M1~\cite{KimACTAMAT2008} (for Ti$_3$C$_2$ and Ti$_4$C$_3$ structures), REAXFF R3~\cite{R3} (for Ti$_2$N structure), and MEAM M2~\cite{KimACTAMAT2008} (for Ti$_3$N$_2$ and Ti$_4$N$_3$ structures).

The four elastic properties of all MXenes were calculated using the selected force fields and then compared with the DFT results found in the literature. 
The linear compressibilities of all titanium MXenes were presented for the first time.
The REAXFF/R4 force field predicts a higher degree of rigidity in titanium carbide MXenes compared to the predictions of DFT calculations. 
The MEAM/M1 force field predicts stiffer structures for the Ti$_3$C$_2$ and Ti$_4$C$_3$ MXenes, with less discrepant values of the elastic properties compared to those from DFT calculation. 

For the titanium nitride MXenes, the REAXFF/R3 force field was selected to describe the properties of only Ti$_2$N MXene. 
While R3 predicts Young's and shear moduli in good agreement with those from DFT calculations, it predicts smaller (larger) values of linear compressibility (Poisson's ratio).
The MEAM/M2 force field predicts softer structures for the Ti$_3$N$_2$ and Ti$_4$N$_3$ MXenes, as compared to DFT calculations. 

A survey of the extant literature reveals that the majority of MD simulations of titanium MXenes without terminations have been performed with the MEAM potential~\cite{MDM4,I43,I47,I49}. 
With the exception of the Ti$_2$C structure, the MEAM force fields yielded highly satisfactory outcomes for both the structure and the elastic properties of all titanium MXenes.
However, certain studies have employed REAXFF force fields, other than the selected R4, such as, for example, R1~\cite{I45} and R2~\cite{I33,I49}. 
Moreover, a number of studies have employed COMB3/C2 force fields~\cite{MDMC1,I48}. 
Provided that the emphasis of these studies is on the structural characteristics of the simulated titanium MXenes, the validity of the research may not be inherently compromised.
As demonstrated in Table \ref{TRSC} and Fig. \ref{fig4}, these force fields provide good or excellent agreement with the structural properties of the corresponding titanium MXenes. 
This and the other findings from the present work indicate that the present study may serve as a valuable resource in selecting the most suitable force field for a specific purpose.

\indent

\bibliographystyle{elsarticle-num} 
\bibliography{mainTiMXenes}

\end{document}